\input epsf
\documentstyle[aps]{revtex}

\begin{document}

\title{Evaporation of alpha particles from $^{31}$P nucleus}

\author{D. Bandyopadhyay,  S. K. Basu, C. Bhattacharya\cite{pa},
S. Bhattacharya, K. Krishan}

\address{ Variable Energy Cyclotron Centre, 1/AF Bidhan Nagar,
Calcutta - 700 064, India}

\author{A. Chatterjee, S. Kailas, A. Navin, A. Srivastava}

\address{Nuclear Physics Division, Bhabha Atomic Research Centre,
Mumbai - 400 085, India}

\maketitle

\begin{abstract}

The   energy   spectra  of  $\alpha$-particles  have  been  measured  in
coincidence with the evaporation residues for the decay of the  compound
nucleus $^{31}$P, produced in the reaction $^{19}$F (96 MeV) + $^{12}$C.
The  data  have  been  compared  with the predictions of the statistical
model code CASCADE. It has been observed  that  significant  deformation
effect in the compound nucleus need to be considered in order to explain
the shape of the evaporated $\alpha$-particle energy spectra.

\end{abstract}


\vspace{0.2 cm}

One  of  the  main  motivations  of  the  low energy heavy ion reactions
studies has been to extract informations on the  statistical  properties
of the hot, rotating nuclei. The informations on the main ingredients of
the statistical description, {\it i.e.,} the nuclear level densities and
the  barrier  transmission  probabilities, are usually obtained from the
study of the evaporated light particle  spectra.  The  validity  of  the
statistical model depends crucially on the successful description of the
light   particle   emission  data  and  the  model,  so  far,  has  been
overwhelmingly successful  in  explaining  a  wide  variety  of  nuclear
reaction  data in low energy regime. In this perspective, recent studies
on the evaporated $\alpha$-particle energy spectra has evoked a  lot  of
interest  (see,  for  example, ref. \cite{1} and references therein). It
has been observed  that  the  standard  statistical  model  calculations
failed  to  predict the shape of the evaporated $\alpha$-particle energy
spectra  satisfactorily.  A  large  number  of  experiments  have   been
performed to study this anomaly over a wide range of compound nuclear
masses A$_{CN}$ in the range of $\sim$60--170 \cite{1,2,2a,3,3a,5},
and  in  all  cases  it  has been found that the average energies of the
measured $\alpha$-particle  energy  spectra  are  much  lower  than  the
corresponding  theoretical  predictions. Several attempts have been made
in the past few years to explain  this  anomaly.  Some  of  the  authors
\cite{2,2a}  argued  that the discrepancy was due to the lowering of the
emission barriers of the hot nuclei as compared to the  fusion  barriers
for the corresponding relatively 'cold' inverse absorption channels as a
result   of   the  excitation  energy  and  angular  momentum  dependent
deformation of the emitting system in the former.  On  the  other  hand,
there is another group of authors who claim that the anomaly may be well
explained  by incorporating spin dependent level density in the standard
statistical model prescription and emission barriers need not be changed
\cite{3,3a}. Moreover, it has also been observed that the  magnitude  of
the  discrepancy  has  some  entrance  channel  dependence \cite{5}, the
discrepancy being more for the more symmetric entrance channels. This is
indicative of the fact that the magnitude of the phenomena may  also  be
linked  with  the  entrance channel dynamics of the system. Intuitively,
the shape of the $\alpha$-particle spectra  would  be  affected  by  the
deformation  of  the  equilibrating system if it remains deformed over a
time scale comparable to the mean life time of  $\alpha$  emission.  For
heavier  nuclei  ($A_{CN}  > 60$), the theoretical calculations of shape
equilibration time, using the code HICOL  \cite{1,5a}  show  that  these
time  scales  are  comparable  to  mean $\alpha$ life times for moderate
excitations. Interestingly, for lighter systems too, the  trend  of  the
predictions are similar. This prompted us to explore the scenario in the
lower mass composite system (A$\leq$60), as there is practically no data
available in this region, to the best of our knowledge.

The  system  chosen for the present studies is $^{31}$P, produced in the
reaction $^{19}$F (96 MeV)  +  $^{12}$C.  The  $\alpha$-particle  energy
spectra  at various laboratory angles, were measured in coincidence with
the evaporation residues. The experiment was  performed  at  the  Bhabha
Atomic  Research  Centre  - Tata Institute of Fundamental Research 14 UD
pelletron accelerator laboratory, Mumbai, with 96 MeV $^{19}$F  beam  on
100$\mu$gm/cm$^2$  self-supporting  $^{12}$C target. Two gas telescopes,
each consisting of a gas $\Delta$E and 500$\mu$m Si surface barrier (SB)
detectors were kept at 20$^\circ$ and 30$^\circ$. The gas detectors were
of axial configuration, continuous flow type and the gas (P10 : 90\% Ar,
10\%  CH$_4$)  pressure   was   maintained   at   80$\pm$1   torr.   The
$\alpha$-particles   have   been   detected   in  coincidence  with  the
evaporation residues in three solid state telescopes in an angular range
of 20$^{\circ}$-60$^{\circ}$. Among these three, two telescopes were  of
two  elements  each  consisting  of  45$\mu$m  $\Delta$E Si(SB) and 2 mm
Si(Li) detectors. The third one was  of  three  elements  consisting  of
30$\mu$m, 46$\mu$m Si(SB) and 2 mm Si(Li) detectors. Gas telescopes were
calibrated using the elastically scattered F ions from C and Bi targets.
The  light  particle  telescopes  were  calibrated  with  recoil protons
produced in the reaction of 96 MeV Flourine on  3.5$\mu$m  mylar  target
and  with  the  radioactive  $^{228}$Th $\alpha$-source. Absolute energy
calibrations were done using standard kinematics and considering  energy
loss calculations.

\begin{figure}[h]
\centering
\epsfxsize=5cm
\epsffile[82 312 545 610]{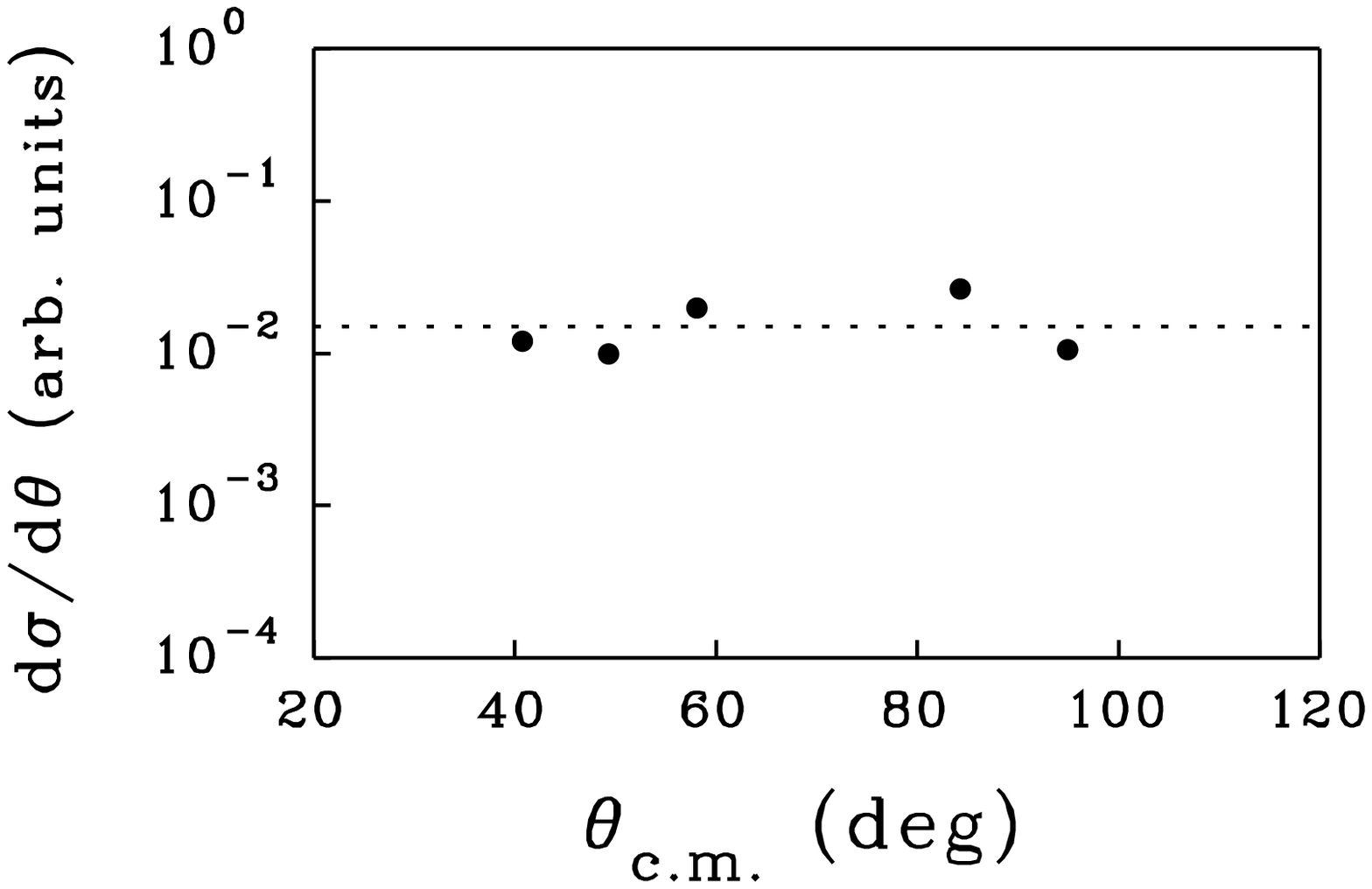}
\caption{Centre-of-mass  angular  distribution, d$\sigma$/d$\theta$. The
dashed   line   correspond   to   fissionlike    angular    distribution
(d$\sigma$/d$\Omega$ $\sim$ 1/sin$\theta$$_{c.m.}$) fits to the data.}

\label{fig1}
\end{figure}

\begin{figure}
\centering
\epsfxsize=5cm
\epsffile[52 290 557 643]{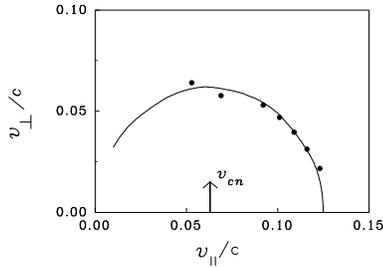}
\caption{Average  velocities  of the $\alpha$-particles as a function of
velocities parallel ($v_{\parallel}$) and perpendicular ($v_{\perp}$) to
the beam direction. The arrow indicates the  velocity  of  the  compound
nucleus.}

\label{fig2}
\end{figure}

The   centre   of   mass   angular   distribution   of   the   exclusive
$\alpha$-particles have been displayed as a function of  the  centre  of
mass  emission angles in Fig.~\ref{fig1}. It is observed from the figure
that the values of d$\sigma$/d$\theta$ is constant over the range of the
observed  centre  of  mass  emission  angles.  This  implies  that   the
d$\sigma$/d$\Omega$     $\sim$    1/sin$\theta$$_{c.m.}$,    which    is
characteristics of the emission from a thermally  equilibrated  compound
nucleus. The average velocities of the exclusive $\alpha$-particles have
been plotted in Fig.~\ref{fig2} as a function of the velocities parallel
($v_{\parallel}$) and perpendicular ($v_{\perp}$) to the beam direction.
It  is  observed that the average velocities fall on a circle around the
compound nucleus velocity ($v_{cn}$)  which  implies  that  the  average
velocities   (as   well   as   the  average  kinetic  energies)  of  the
$\alpha$-particles are  independent  of  the  centre  of  mass  emission
angles.  It  is  a  further  indication  of  the  fact  that  the energy
relaxation is complete and $\alpha$-particles are emitted from  a  fully
equilibrated source moving with the velocity $v_{cn}$.

\begin{figure}
\centering
\epsfxsize=10cm
\epsffile[51 221 509 762]{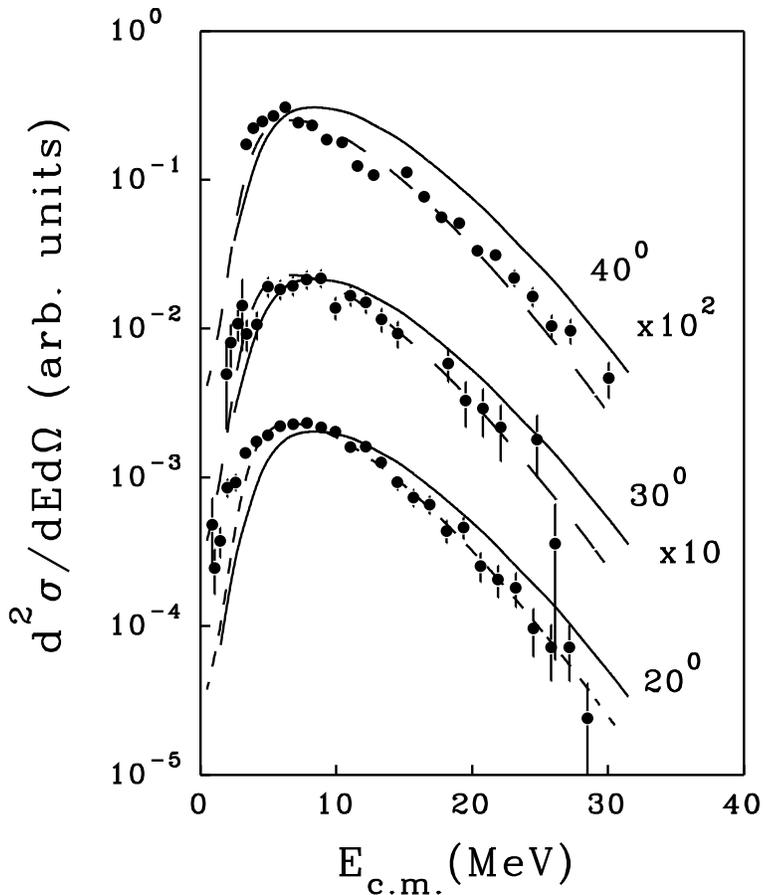}

\caption{Exclusive  (filled circles) $\alpha$-particle energy spectra in
centre-of-mass frame measured at different laboratory angles. Curves are
the prediction of CASCADE calculations with the default  (solid  curves)
and  the  modified (dashed curves) values of the radius parameter, $r_0$
({\it see text}).}

\label{fig3}
\end{figure}

The  exclusive  c.m.  energy  spectra  of  $\alpha$-particle  for  three
representative  angles  have  been   compared   with   the   theoretical
predictions of the same using the code CASCADE \cite{6}, and the results
are  displayed  in  Fig.~\ref{fig3}.  For  the  present calculation, the
critical  angular  momentum  for  fusion,  $l_{cr}$,  was  taken  to  be
21$\hbar$,  obtained  from  the compilation of fusion cross section data
\cite{7}. The solid curves are the results of CASCADE calculations  with
the default parameters of the code. It is clear from the figure that the
theoretical  predictions  using  the  default  values  of the parameters
differ significantly from the experimental data on both lower and higher
energy sides of the spectra. It is  also  clear  that  the  experimental
values  of  the mean $\alpha$- particle energies are somewhat lower than
those predicted from the theory. It  implies  that  some  of  the  input
parameters of the code need to be modified to explain the data properly.

The  change  in  the  emission  barriers  and vis-a-vis the transmission
probabilitiies  affects  the  lower  energy  part  of   the   calculated
evaporation  spectra.  On  the  other  hand  the high energy side of the
spectra depends crucially on the available phase space obtained from the
energy level densities. In hot  rotating  nuclei  formed  in  heavy  ion
reactions,  the  energy level density at higher angular momentum is spin
dependent. The level density, $\rho(E,I)$, for a given angular  momentum
$I$ and excitation energy $E$ is given by \cite{6},

\begin{equation}
\rho (E,I) = ((2I+1)/12)a^{1/2} (\hbar^2/2 {\cal J}_{eff} )^{3/2}\\
		 1/(E-\Delta-T-E_I)^2 exp[2[a(E-\Delta-T-E_I)]^{1/2}]
\label{lev}
\end{equation}

where $a$ is the level density parameter, which is taken to be $A/8$ for
the  present calculation, $T$ is the thermodynamic temperature, $\Delta$
is the pairing correction, and E$_I$ is the rotational energy which  can
be  written in terms of the effective moment of inertia ${\cal J}_{eff}$
as

\begin{mathletters}
\begin{equation}
E_I =\frac{ \hbar^2}{2 {\cal J}_{eff}}I(I+1),
\end{equation}
where
\begin{equation}
{\cal J}_{eff}= {\cal J}_0 \times (1+\delta_1I^2+\delta_2I^4),
\label{jeff}
\end{equation}

and the rigid body moment of inertia,  ${\cal J}_0$, is given by,

\begin{equation}
{\cal J}_0 = \frac{2}{5} A^{5/3} r_{\circ}^2.
\end{equation}
\end{mathletters}

Here,  A  is  the mass number and r$_\circ$ is the radius parameter. Non
zero values of the  parameters  $\delta_1$,  $\delta_2$  introduce  spin
dependence  in  the  effective  moment  of  inertia  resulting  in  spin
dependent level densities in Eq. \ref{lev}.

An  increase  in  the  radius parameter $r_0$ results in an increase the
available phase space, due to an increase in  the  effective  moment  of
inertia.  Simultaneously, the transmission probability also increases as
the emission barrier decreases due to the increase of $r_0$.  Therefore,
with  the  variation  of  a  single  input parameter, $r_0$, in the code
CASCADE, it is possible to modify the low energy as  well  as  the  high
energy  part  of the calculated spectra. So, in the present calculation,
we varied  the  parameter  $r_0$  only,  and  kept  the  spin  dependent
parameters $\delta_1$, $\delta_2$ in Eqn.~(\ref{jeff}) equal to zero, to
reproduce the experimental spectra. The dashed curves in Fig.~\ref{fig3}
are  the  predicted energy spectra using the code CASCADE with the value
of $r_0=1.56$ fm. It is clear from Fig.~\ref{fig3} that  the  calculated
spectra agree quite well with the exclusive experimental spectra both in
the  low  energy  as well in the high energy regions. The large value of
the parameter $r_0$ (as compared  to  its  default  value  of  1.29  fm)
indicates that the nucleus is deformed in the exit channel. The increase
in the radius parameter $r_0$ is $\sim 20\%$ which also agrees well with
the   other   observations   available   in   the  literature  \cite{1}.
Interestingly, similar deformation effects have  been  observed  in  our
earlier  studies  on  intermediate mass fragment (IMF) emission from the
same $^{31}$P nucleus \cite{8}.

In  summary,  we  have  measured  $\alpha$-particle  energy  spectra  in
coincidence with the evaporation residues.  The  $\alpha$-particles  are
emitted  from  an  equilibrated  compound  nucleus as evidenced from the
angular distributions and the velocity  diagrams.  The  measured  energy
spectra  are  underpredicted  by the statistical model calculations with
spherical  compound  nuclear  configuration.  However,  a   satisfactory
description of the measured energy spectra has been achieved by invoking
a   deformed   configuration   of   the  compound  nucleus  through  the
modification of the radius parameter $r_0$. Thus, it  may  be  concluded
that  the  evaporation spectra of $\alpha$-particles from light compound
nuclear systems, such as $^{31}P$, can be satisfactorily  reproduced  by
introducing  only  an  effective deformation in the exit channel through
the modification of the radius parameter.  Further  experiments  may  be
needed  to  confirm  this  finding in other light nuclear systems in the
range $A \leq 60$.

\acknowledgements

The  authors  thank  the Pelletron operating staff for smooth running of
the machine and Mr. D. C. Ephraim of T.I.F.R., for making the targets.

\pagebreak

\end{document}